# The Impact of Sampling and Rule Set Size on Generated Fuzzy Inference System Predictive Accuracy: Analysis of a Software Engineering Data Set

Stephen G. MacDonell

*SERL, Auckland University of Technology
Private Bag 92006, Auckland 1142, New Zealand*
stephen.macdonell@aut.ac.nz

**Abstract**

*Software project management makes extensive use of predictive modeling to estimate product size, defect proneness and development effort. Although uncertainty is acknowledged in these tasks, fuzzy inference systems, designed to cope well with uncertainty, have received only limited attention in the software engineering domain. In this study we empirically investigate the impact of two choices on the predictive accuracy of generated fuzzy inference systems when applied to a software engineering data set: sampling of observations for training and testing; and the size of the rule set generated using fuzzy c-means clustering. Over ten samples we found no consistent pattern of predictive performance given certain rule set size. We did find, however, that a rule set compiled from multiple samples generally resulted in more accurate predictions than single sample rule sets. More generally, the results provide further evidence of the sensitivity of empirical analysis outcomes to specific model-building decisions.*

**Keywords**: fuzzy inference, prediction, software size, source code, sampling, sensitivity analysis

## 1. INTRODUCTION

Accurate and robust prediction of software process and product attributes is needed if we are to consistently deliver against project management objectives. In recent years fuzzy inference systems (FIS) have gained a degree of traction (with empirical software engineering researchers if not practitioners) as an alternative or complementary method that can be used in the prediction of these attributes [1-4].

Empirical software engineering research that has focused on prediction is normally performed using a standard sample-based approach in which a data set is split into two sub-samples, the first, larger set for building a predictive model and the second for non-biased evaluation of predictive performance (for instance, see [1]). Any reasonable splitting strategy can be used. Typically, two-thirds of the total sample is randomly allocated to the build subset, leaving one third of the observations to comprise the test subset. Given the incidence of outlier observations and skewed distributions in software engineering data sets [5], stratified sampling may be useful, but it has not been widely employed in software engineering analyses. Similarly, where information on the ordering of observations is available this could be used to further inform the sampling strategy. Again, however, such an approach has not been used to any great extent in empirical software engineering research [6].

In this paper we analyze the sensitivity of prediction outcomes to systematic changes in two parameters relevant to fuzzy inference model building and testing: data sampling and rule set size. In essence, then, this is a paper focused on the *infrastructure* of predictive modeling – our intent is to highlight the potential pitfalls that can arise if care is not taken in the building and testing of models. In the next section we describe the empirical work we have undertaken using a software engineering data set. Section 3 reports the results of our analysis. A discussion of the key outcomes is provided in Section 4, after which we briefly address related work. We conclude the paper and highlight opportunities for future research in Section 6.

## 2. EMPIRICL ANALYSIS

In this section we describe the empirical analysis undertaken to assess the impact of changes in both data sampling and rule set size on the predictive accuracy of fuzzy inference systems. After providing a description of the data set employed in the analysis we explain the three approaches used in constructing our prediction systems.

## 2.1. The Data Set

The data set we have used here is a simple one in that it involves a small number of variables. This does not detract, however, from its effectiveness in illustrating the impact of sample selection and rule set size, the principal issues of concern in this study. The data were collected in order to build prediction systems for software product size. The independent variables characterized various aspects of software specifications, including the number of entities and attributes in the system data model, the number of data entry and edit screens the system was to include, and the number of batch processes to be executed. As each implemented product was to be built using a 4GL (PowerHouse) our dependent variable was the number of non-comment lines of 4GL source code. The systems, which provided business transaction processing and reporting functionality, were all built by groups of final-year undergraduate students completing a computing degree.

A total of 70 observations were available for analysis. Although this is not a large number, it did represent the entire population of systems built in that environment over a period of five years. It is in fact quite a homogeneous data set – although the systems were built by different personnel over the five-year period they were all constructed using the same methodology and tool support by groups of four people of similar ability. Furthermore, it is quite common in software engineering to have access to relatively small numbers of observations (for example, see [7]), so the data set used here is not atypical of those encountered in the domain of interest.

In a previous investigation [2] we identified that two of the 8 independent variables were significantly correlated to source code size and were also not correlated to one another. These were the total number of attributes in the system data model (Attrib) and the number of non-menu functions (i.e. data entry and reporting modules rather than menu selection modules) in the system's functional hierarchy (Nonmenu). We therefore used these two variables as our predictors of size.

## 2.2. FIS Development

We adopted three model-building approaches in order to investigate the impact of sampling and fuzzy rule set size on the coverage and accuracy of source code size predictions. We first built membership functions and rule sets (via the algorithms below) using the entire data set of 70 observations (Full Approach); we also randomly generated ten separate build subsets of 50 cases that we used to build fuzzy prediction systems (Sampled Approach); finally we analyzed the ten rule sets generated by the sampling process to create a 'mega' rule set comprising the 50 most frequently generated rules (Top 50 Approach). For each of the three approaches we generated fuzzy inference systems comprising from one to fifty rules. Each FIS was tested against the ten test subsets of 20 observations. While performance was assessed using a variety of accuracy indicators, we report here the results in terms of the sum of the absolute residuals and the average and median residual values, as these are preferred over other candidate measures [8]. We also assessed the coverage of each FIS i.e. the proportion of the test subsets for which predictions were generated (meaning the rule set contained rules relevant to the test subset observations that were then fired to produce a predicted size value).

Previous research using this data set [2] had indicated that the three variables of interest were most effectively represented by seven triangular membership functions, mapping to the concepts <VerySmall, Small, SmallMedium, Medium, MediumLarge, Large, VeryLarge>. The membership function extraction algorithm is as follows:

1. select an appropriate mathematically defined function for the membership functions of the variable of interest ($i$), say $f_i(x)$

2. select the number of membership functions that are desired for that particular variable, $m_i$ functions for variable $i$ ($m_i$ may be set by the user or may be found automatically, to desired levels of granularity and interpretability)

3. call each of the $m_i$ functions $f_{ij}([x])$ where $j = 1\ldots m_i$ and $[x]$ is an array of parameters defining that particular function (usually a center and width parameter are defined, either explicitly or implicitly)

4. using one-dimensional fuzzy c-means clustering on the data set find the $m_i$ cluster centers, $c_{ij}$ from the available data ($m_i$ may be set by the user or may be found automatically, to desired levels of granularity and interpretability)

5. sort the cluster centers $c_{ij}$ into monotonic (generally ascending) order for the given $i$

6. set the membership function center for $f_{ij}$, generally represented as one of the parameters in the array $[x]$, to the cluster center $c_{ij}$

7. set the membership function widths for $f_{ij}$ in $[x]$ such that $\sum_{n=1}^{m_i} f_{in}([c_{in},\ldots]) = 1$, or as close as possible for the chosen $f(x)$ where this cannot be achieved exactly (for example for triangular membership functions each function can be defined using three points, $a$, $b$, and $c$ where $a$ is the center of the next smaller function and $c$ is the center of the next larger function).

Rules were extracted using the same clustering process with multiple dimensions (matching the number of antecedents plus the single consequent):

1. start with known membership functions $f_{ij}([x])$ for all variables, both input and output, where $j$ represents the number of functions for variable $i$ and $[x]$ is the set of parameters for the particular family of function curves

2. select the number of clusters $k$ (which represents the number of rules involving the $k$-1 independent variables to estimate the single output variable)

3. perform fuzzy c-means clustering to find the centers ($i$ dimensional) for each of the $k$ clusters

4. for each cluster $k$ with center $c_k$

    (a) determine the $k^{th}$ rule to have the antecedents and consequent $f_{ij}$ for each variable $i$ where $f_{ij}$ ($c_k$) is maximized over all $j$

    (b) weight the rule, possibly as $\prod_{n=1}^{i} f_{ij}(c_k)$ or $\sum_{n=1}^{i} f_{ij}(c_k)$

5. combine rules with same antecedents and consequents, either summing, multiplying, or bounded summing rule weights together

6. (optionally) ratio scale all weights so that the mean weight is equal to 1.0 to aid interpretability.

## 3. RESULTS

We now consider the outcomes of each of the three approaches in turn – Full, Sampled and Top 50.

### 3.1. Full Approach

Development of an FIS using the entire set of observations is the most optimistic of the three approaches, and in fact represents more of a model-fitting approach rather than one of unbiased prediction where a hold-out sample is used. We employed this approach, however, to provide a benchmark for comparative model performance. The clustering approaches described above were used to build membership functions and rule sets based on all 70 observations. Essentially this meant that there were fifty distinct FIS produced, comprising from one to fifty rules. These FIS were then tested against each of the ten test subsets produced from the Sampled Approach (described below). Accuracy and coverage were assessed for each FIS, illustrated by the example shown for the sample 4 test subset in Figures 1 and 2 and Table 1.

For sample 4 we can see in Figure 1 that predictive accuracy measured using the median residual value taken over the predictions made shows some volatility but is generally stable within the range 150-250 SLOC, once the rule set size reaches a value of around 18. At this point also we can see that coverage becomes stable at close to 100% i.e. rules were fired and predictions made for close to all 20 test subset observations. Further evidence of the generally high degree of test observation coverage is provided in Table 1 and Figure 2. This latter result is not unexpected as in the Full Approach all 70 observations were used to build the FIS, and so the test observations had been 'seen' and accommodated during model-building.

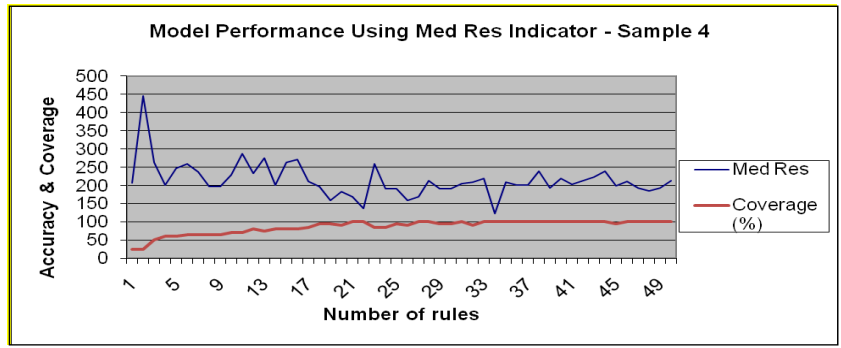

**Fig. 1.** Change in median residual and coverage with increasing rule set size for sample 4.

**Table 1.** Overall coverage for sample 4.

| Coverage | Sample 4 | | | |
|---|---|---|---|---|
| | | | | |
| 100% | 22 | 44% | Mean | 86% |
| 80%-99% | 16 | 32% | Med | 95% |
| 60%-79% | 9 | 18% | | |
| 40%-59% | 1 | 2% | | |
| 20%-39% | 2 | 4% | | |
| 0%-19% | 0 | 0% | | |

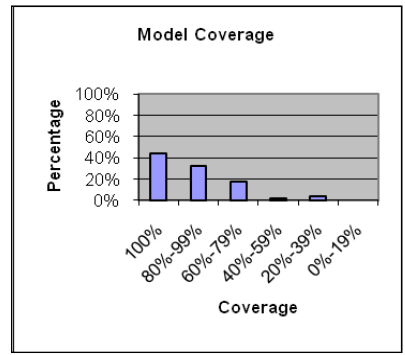

**Fig. 2.** Graphical depiction of coverage.

## 3.2. Sampled Approach

As stated above, in this approach we randomly allocated 50 of the 70 observations to a build subset and left the remaining 20 observations for testing, repeating this process ten times to create ten samples. FIS development then used only the build subsets, creating FIS comprising an increasing number of rules (from one to fifty, the latter meaning one rule per build observation) for each of the ten samples.

**Sampling.** We performed t-tests to assess the degree of uniformity across the ten samples, using Size as our variable of interest. These tests revealed that, in particular, sample 10 was significantly different (at alpha=0.05) from samples 3 through 8. This could also be verified informally by looking at the data – build subset 10 comprised a higher proportion of larger systems (and test subset 10 a correspondingly higher proportion of smaller systems) than most of the other samples. As our objective was to illustrate the impact of such sampling on model performance and the rule sets generated, however, we retained the samples as originally produced.

**Model performance.** Model accuracy varied across the ten samples, and while in general there was a tendency for accuracy to either stabilize or improve as the FIS rule set grew (as per Figure 1 above for the Full Approach) this was not found to always be the case. For instance, prediction of the test subset for sample 2 was quite volatile in spite of good coverage, as shown in Figure 3.

The best models achieved for each sample are shown in Tables 2 and 3. The first table depicts the results irrespective of coverage whereas the second considers only those FIS that achieved maximum coverage for each test subset. What is evident in both tables is the variation in values for all three accuracy measures – for instance, the median residual value in the best maximum coverage models (Table 3) varies from a low of 139 SLOC (for sample 2) up to 296 SLOC (for sample 8).

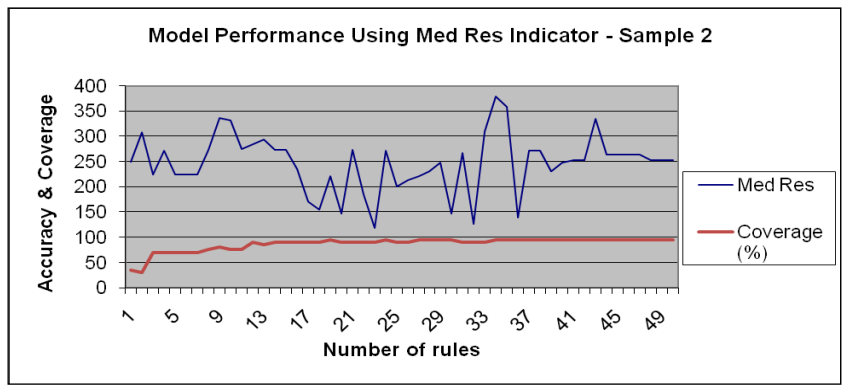

**Fig. 3.** Change in median residual and coverage with increasing rule set size for sample 2.

**Table 2.** Best models for each sample in terms of accuracy measures (when coverage ignored).

| Sample | 1 | 2 | 3 | 4 | 5 | 6 | 7 | 8 | 9 | 10 |
|---|---|---|---|---|---|---|---|---|---|---|
| Coverage | 80% | 90% | 90% | 100% | 90% | 95% | 80% | 90% | 85% | 95% |
| Minimum rules? | 19 | 23 | 31 | 34 | 18 | 14 | 12 | 35 | 29 | 26 |
| Abs Res | 3100 | 4340 | 5430 | 6156 | 4077 | 4623 | 5257 | 5996 | 6823 | 5000 |
| Ave Res | 194 | 241 | 302 | 308 | 227 | 243 | 329 | 333 | 401 | 263 |
| Med Res | 111 | 118 | 163 | 177 | 159 | 200 | 266 | 288 | 261 | 176 |

**Table 3.** Best models for each sample in terms of accuracy measures (maximum coverage).

| Sample | 1 | 2 | 3 | 4 | 5 | 6 | 7 | 8 | 9 | 10 |
|---|---|---|---|---|---|---|---|---|---|---|
| Coverage | 95% | 95% | 100% | 100% | 100% | 100% | 95% | 100% | 100% | 100% |
| Minimum rules? | 44 | 36 | 40 | 34 | 50 | 15 | 38 | 49 | 45 | 36 |
| Abs Res | 5825 | 5017 | 6394 | 6156 | 7193 | 4807 | 5970 | 5842 | 6789 | 5028 |
| Ave Res | 307 | 264 | 320 | 308 | 360 | 240 | 314 | 292 | 339 | 251 |
| Med Res | 187 | 139 | 208 | 177 | 274 | 203 | 177 | 296 | 269 | 173 |

**Rule Distribution.** For each of the ten samples we also considered the composition of the generated rule sets, comprising from one to fifty rules each at increments of one, meaning a total of 1275 rules per sample. Each rule was a triple made up of two antecedents and a single consequent e.g. IF <Attrib> IS [Small] AND <Nonmenu> IS [SmallMedium] THEN <Size> IS [Small]. Our analysis considered the frequency of particular memberships in the rule sets for each variable (e.g. count of '<Attrib> IS [VerySmall]'), as well as the frequency of each complete rule across the sets. Space precludes full reporting of the results of this analysis across all ten samples but an example for sample 6 is shown in Figures 4, 5 and 6. (Note that in these figures Set Value '1' maps to 'VerySmall', '2' maps to 'Small', and so on.)

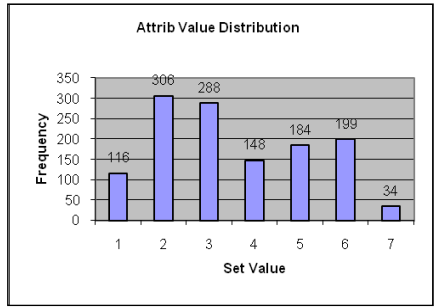

**Fig. 4.** Membership frequency for Attrib.

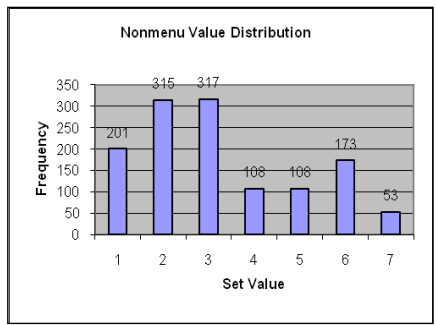

**Fig. 5.** Membership frequency for Nonmenu.

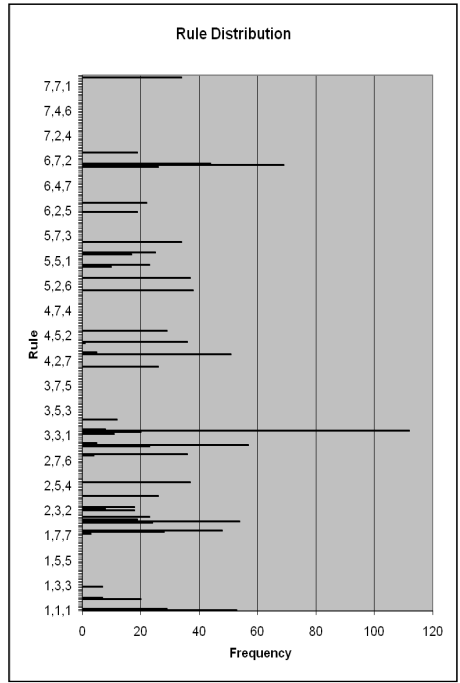

**Fig. 6.** Rule frequency for sample 6.

For this sample 47 distinct rules of the 343 possible (the latter representing each possible triple from '1,1,1' through to '7,7,7') made up the 1275 rules in the fifty sets, the most frequently occurring rule for this sample being '3,3,4': IF <Attrib> IS [SmallMedium] AND <Nonmenu> IS [SmallMedium] THEN <Size> IS [Medium], with 112 instances over the fifty rule sets (shown as the longest bar in Figure 6). Across the ten samples, the number of distinct rules varied from 45 for sample 8 up to 60 for sample 1, and these sets were distinct in part, meaning that over the ten samples 131 distinct rules were generated from the 343 possible candidates.

### 3.3. Top 50 Approach

The above analysis of the ten samples allowed us to also identify the Top 50 distinct rules *across* the sampled data sets i.e. the 50 rules that were generated most frequently (where 50 was an arbitrary choice). The most common rule (at 506 instances) was: IF <Attrib> IS [MediumLarge] AND <Nonmenu> IS [MediumLarge] THEN <Size> IS [MediumLarge], whereas the 50[th] most common rule, with 89 instances, was IF <Attrib> IS [SmallMedium] AND <Nonmenu> IS [Small] THEN <Size> IS [VerySmall]. We then applied this set of 50 rules to the ten test subsets, in the expectation that over the entire set they may outperform the potentially over-fitted Sampled rule sets generated from each specific build subset. We compare the results obtained from this approach to those derived from the originally Sampled approach.

### 3.4. Comparison of Approaches

As might be expected, the Full approach models performed well across the ten samples. When considered in terms of the sum of absolute residual and average and median residual measures, the Full model was the most accurate in 14 of the 30 cases (30 being 10 samples by 3 error measures). As noted above, however, this does not represent an unbiased test of predictive accuracy as the Full models were built using the entire data set. Our comparison is therefore focused on the Sampled and Top 50 approaches. Figure 7 illustrates one such comparison, based on the median residual measure for sample 3. We can see that in this case the model created from the Top 50 approach performed better than those from the Sampled approach for FIS comprising up to 29 rules, but beyond that the Sampled approach was generally superior.

The pattern exhibited for sample 3 was not a common one across the ten samples, however. The data presented in Tables 4 and 5 indicate that, in general, the Top 50 model ('Top') outperformed the specific FIS developed with the Sampled approach ('Sam.'). Of the 30 comparisons, 9 favored the Sampled approach. (Note that totals can exceed 100% for a sample because the two methods may achieve equivalent performance for a given number of rules, and so are both considered 'Best'.)

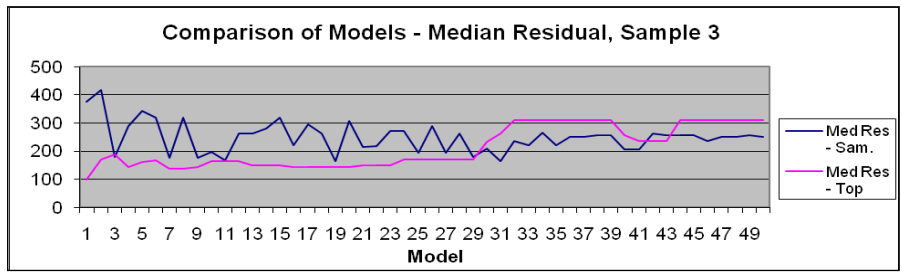

**Fig. 7.** Sample 3 comparison in terms of median residual (Sam.: Sampled, Top: Top 50).

**Table 4**. Comparison of model accuracy across the ten samples (best shown in bold type).

| Sample | 1 | | 2 | | 3 | | 4 | | 5 | |
|---|---|---|---|---|---|---|---|---|---|---|
| Approach | Sam. | Top | Sam. | Top | Sam. | Top | Sam. | Top | Sam. | Top |
| Abs Res Best | **52%** | 50% | 50% | **52%** | **52%** | 48% | 14% | **86%** | 14% | **86%** |
| Ave Res Best | 38% | **64%** | 24% | **78%** | 36% | **64%** | 10% | **90%** | 8% | **92%** |
| Med Res Best | 42% | **60%** | **56%** | 46% | 40% | **60%** | 30% | **72%** | 0% | **100%** |
| Sample | 6 | | 7 | | 8 | | 9 | | 10 | |
| Approach | Sam. | Top | Sam. | Top | Sam. | Top | Sam. | Top | Sam. | Top |
| Abs Res Best | 16% | **84%** | **82%** | 20% | 26% | **74%** | 30% | **72%** | **70%** | 30% |
| Ave Res Best | 24% | **76%** | **74%** | 28% | 6% | **94%** | 20% | **82%** | **62%** | 38% |
| Med Res Best | 20% | **82%** | **58%** | 46% | 12% | **88%** | 22% | **80%** | **72%** | 28% |

**Table 5**. Comparison of model accuracy – summary statistics.

| Summary Statistic | Mean | | Median | |
|---|---|---|---|---|
| Approach | Sam. | Top | Sam. | Top |
| Abs Res Best | 41% | 60% | 40% | 62% |
| Ave Res Best | 30% | 71% | 24% | 77% |
| Med Res Best | 35% | 66% | 35% | 66% |

## 4. DISCUSSION

The impact of sampling was evident in the t-tests of difference in Size across the ten samples. The mean value for Size varied from 1043 to 1211 SLOC in the build subsets, and from 843 to 1264 in the test subsets, with a significant difference evident in relation to sample ten. Similarly, FIS model performance also varied over the samples in terms of both the residual values and the coverage of the test observations. For example, the outright best model for sample 1 comprised just 19 rules, had an absolute error value of 3100 SLOC and achieved 80% coverage, whereas the sample 4 best model achieved 100% coverage using 34 rules and an absolute residual of 6156 SLOC. No consistent pattern was evident in relation to rule set size even for FIS achieving maximum coverage – across the ten samples the most accurate FIS was obtained from rule sets varying in size from 15 to 49 rules (out of a possible 50 rules). The rules also varied from one sample to the next (aligned with the differences in the data) – 131 distinct rules were generated. These outcomes all reinforce the importance of considering multiple samples and aggregating results in cases where there is variance in the data set, rather than relying on a single or small number of splits.

The Top 50 rules (in terms of frequency) generally outperformed the single-sample FIS, struggling only against samples 7 and 10. Overall this is an expected result, as the mega-set of rules was derived from exposure to most if not all 70 observations, albeit across ten samples, and we were already aware that sample 10 was significantly different to the others. The poor performance on sample 7 needs further analysis.

## 5. RELATED WORK

As far as we are aware there have been no prior *systematic* assessments of the impact of these parameters on analysis outcomes in the empirical software engineering community. While several studies have used FIS in software engineering (e.g. [3]) only single splits of the data set have been used, and there has been no specific assessment of the impact of rule set size. That said, [4] describes the selection of rules based on error thresholds. Related work has also been undertaken in other domains. For instance, [9] considered the trade-off between a smaller rule set and the accuracy of decisions made in regard to credit scoring, concluding that substantially fewer rules did not lead to proportional reductions in accuracy. The author notes, however, that "[t]he cost in accuracy loss one is willing to pay for the benefit of a smaller rule-base is entirely domain and organization dependent." [9, p.2787]. [10] investigated the impact of stratification of training set data in data mining (employing evolutionary algorithms) noting significant differences in classification outcomes. Within the software engineering research community the influence of sampling on data analysis has received some attention, but as this does not relate to fuzzy inference the interested reader is referred to [11] for an example of this work.

## 6. CONCLUSIONS AND FUTURE WORK

We have conducted an analysis of the sensitivity of predictive modeling accuracy to changes in sampling and rule set size in the building and testing of fuzzy inference systems for software source code size. Our results indicate that both aspects influence analysis outcomes – in particular, the splits of data across build and test sets lead to significant differences in both predictive accuracy and test set coverage.

We are continuing to work on aspects of prediction infrastructure – in particular, our current work is focused on considering the impact of temporal structure on modeling outcomes as well as the development of more suitable error measures for accuracy assessment. Further research should also address the degree to which stability in predictor variables over the life of a project affects predictions of size, quality and other aspects of project management interest.

## REFERENCES

1. Azzeh, M., Neagu, D., Cowling, P.: Improving Analogy Software Effort Estimation Using Fuzzy Feature Subset Selection Algorithm. In: 4th Intl Workshop on Predictive Models in Softw. Eng., pp. 71--78. ACM Press (2008)

2. MacDonell, S.G.: Software Source Code Sizing Using Fuzzy Logic Modeling. Info. and Softw. Tech. 45(7), 389--404 (2003)

3. Huang, X., Ho, D., Ren, J., Capretz, L.F.: A Soft Computing Framework for Software Effort Estimation. Soft Computing 10, 170--177 (2006)

4. Ahmed, M.A., Saliu, M.O., AlGhambi, J.: Adaptive Fuzzy Logic-based Framework for Software Development Effort Prediction. Info. and Softw. Tech. 47, 31--48 (2005)

5. Kitchenham, B., Mendes, E.: Why Comparative Effort Prediction Studies may be Invalid. In: 5th Intl Conf. on Predictor Models in Softw. Eng., pp. in ACM DL. ACM Press (2009)

6. MacDonell, S.G., Shepperd, M.: Data Accumulation and Software Effort Prediction. In: 4th Intl Symp. on Empirical Softw. Eng. and Measurement, pp. in ACM DL. ACM Press (2010)

7. Song, Q., Shepperd, M.: Predicting Software Project Effort: A Grey Relational Analysis Based Method. Expert Systems with Applications 38, 7302--7316 (2011)

8. Kitchenham, B.A., Pickard, L.M., MacDonell, S.G., Shepperd, M.J.: What Accuracy Statistics Really Measure. IEE Proc. - Software 148(3), 81--85 (2001)

9. Ben-David, A.: Rule Effectiveness in Rule-based Systems: A Credit Scoring Case Study. Expert Systems with Applications 34, 2783--2788 (2008)

10. Cano, J.R., Herrera. F., Lozano, M.: On the Combination of Evolutionary Algorithms and Stratified Strategies for Training Set Selection in Data Mining, Applied Soft Computing 6, 323--332 (2006)

11. Shepperd, M., Kadoda, G.: Using Simulation to Evaluate Prediction Techniques. In: 7th Intl Symp. on Softw. Metrics, pp. 349--359, IEEE CS Press (2001)